# Tailoring femtosecond hot-electron pulses for ultrafast spin manipulation

N. Bergeard[1,2], M. Hehn[1], K. Carva[3], P. Baláž[3], S. Mangin[1], G. Malinowski[1]

[1] Institut Jean Lamour, CNRS - Université de Lorraine, 54011 Nancy, France.

[2] Université de Strasbourg, CNRS, Institut de Physique et Chimie des Matériaux de Strasbourg, UMR 7504, F-67000 Strasbourg, France.

[3] Department of Condensed Matter Physics, Faculty of Mathematics and Physics, Charles University, Ke Karlovu 5, CZ-121 16 Prague, Czechia

## *0.* Abstract

We have measured the hot-electron induced demagnetization of a [Co/Pt]$_2$ multilayer in M(x nm)/Cu(100 nm)/[Co(0.6 nm)/Pt(1.1 nm)]$_2$ samples depending on the nature of the capping layer M and its thickness x. We found out that a Pt layer is more efficient than [Co/Pt]$_X$, Cu or MgO layers in converting IR photon pulses into hot-electron pulses at a given laser power. We also found out that the maximum relative demagnetization amplitude is reached for M(x) = Pt (7 nm). Our experimental results show qualitative agreement with numerical simulations based on the superdiffusive spin transport model. We concluded that the maximum relative demagnetization amplitude, which corresponds to the highest photon conversion into hot-electrons, is an interplay between the IR penetration depth and the hot-electron inelastic mean free path within the capping layer.

## *1.* Introduction

The perspective of ultrafast magnetization manipulation by ultrashort stimuli is a promising route towards faster data processing in magnetic storage devices [1]. Thermally induced magnetization switching in FeCoGd alloys by a single femtosecond (fs) infrared (IR) laser pulse with linear polarization [2] is the fastest determinist process for magnetization manipulation ever reported so far [3 - 5]. The microscopic origin of this phenomenon, usually labelled AO-HIS for All Optical Helicity-Independent Switching, is still heavily debated in spite of intensive theoretical researches [6 - 11]. Recently, Lalieu et al. have shown that AO-HIS is not limited to FeCoGd ferrimagnetic alloy by evidencing single-shot magnetization reversal in a Gd/Co/Pt stacks [12, 13]. Such a discovery suggests that it may be possible to observe AO-HIS in technological relevant ferrimagnetic materials by tailoring their magnetic properties. In the meanwhile, thermally induced magnetization reversal on the picosecond time-scale by a single unpolarized electron pulse was also demonstrated in FeCoGd ferrimagnetic alloy [14] which allows envisioning integration in modern data storage devices based on manipulation of magnetization by electric currents as demonstrated by Yang et al [15]. Xu *et al.* have even shown that the magnetization reversal by a single femtosecond hot-electron pulse can be as fast as the light induced reversal [16]. Such femtosecond hot-electron pulses can be generated by photo-exciting a thick non-magnetic capping layer deposited on top of the magnetic material [17 - 20]. The manipulation of magnetization on the picosecond time scale is a lever for improving the performances of current data storage devices. Finding the most appropriate non-magnetic capping layer to convert the IR fs laser pulse into a hot-electron pulse is thus of paramount importance in this optics. Therefore, in this work, we aimed at evaluating and comparing the conversion efficiency (CE) of various materials and thicknesses as capping layer for the generation of hot-electron pulses. As recently demonstrated, this evaluation can be done by monitoring the hot-electron induced demagnetization of a buried magnetic layer [21, 22].

The transport of hot electrons in metals and their role in magnetization dynamics can be described by the superdiffusive spin-dependent transport model [23, 24], which is based on semi classical equations of motion for electrons moving in two spin channels in a certain range of energies above the Fermi level. Assuming given initial distribution of photo-excited electrons, the electronic transport is governed by the material, energy and spin-dependent electron velocities and life times. The asymmetry of the transport properties with respect to the electron spin then leads to the ultrafast demagnetization in ferromagnets as well as the spin transfer torque in non-collinear magnetic multilayers, which has been successfully described within this model [25].

In this work, we have monitored the relative amplitude of the hot-electron induced demagnetization (**q**) in [Co/Pt] multilayers [17 - 20, 26, 27] for various capping layers. We found a qualitative agreement between our experimental data and the numerical simulations based on the superdiffusive spin transport model [23, 24]. It shows that the maximum q, which correspond to the highest CE, is an interplay between the IR penetration depth and the hot-electron inelastic mean free path within the capping layer.

*2.* **Materials and methods**

The Glass/Ta(3)/Pt(3)/[Co(0.6)/Pt(1.1)]2/Co(0.6)/Cu(100)/M(x) multilayers with M(x)= Pt(3) (sample 1), [Co(0.6)/Pt(1.1)]$_4$ (sample 2), MgO(5) (sample 3) or no capping layer (sample 4) were grown by dc magnetron sputtering under ultrahigh vacuum. We have also grown additional Glass/Ta(3)/Pt(3)/[Co(0.6)/Pt(1.1)]2/Co(0.6)/Cu(100)/Pt(x) multilayers. In these samples, we varied the Pt thicknesses from x=3 to 20. All thicknesses in brackets are given in nanometers. The various samples investigated were deposited subsequently which ensures identical magnetic properties for the [Co(0.6)/Pt(1.1)]2/Co(0.6)] (Co/Pt). In all cases, the Co/Pt multilayer has an out of plane anisotropy. The Cu(100) thickness ensures that the Co/Pt was not directly excited by the fs-laser pulses but by the photoexcited hot-electrons [19].

We have used two experimental excitation schemes to carry out our investigations: the "pump-front / probe-back" and "pump-back / probe-back" configurations (figure 1). In the former configuration, hot-electron pulses were generated by exciting the top layer with an 800 nm laser pulse while the hot-electron induced demagnetization was probed by mean of time-resolved Magneto-Optical Kerr Effect from the back side of the sample using a 400 nm probe. In the latter configuration, both the pump and probe pulses hit the Co/Pt from the back side of the multilayers. This configuration allows recording the direct laser induced demagnetization to get a correlation between the Kerr rotation in the excited state and the actual demagnetization amplitude. In both configurations, the pump and probe pulses arrive on the sample at normal incidence. The pump diameter was set to approximately to 300 µm while the probe diameter was set to 60 µm ensuring homogeneous excitations over the probed area of the samples. Further details on the configurations, the laser set-up and normalization can be found in [19] and its supplement materials.

### *3.* Results and discussion

In figure 2a, we report the transient Kerr rotation upon hot-electron induced excitation as a function of the pump-probe delay for the samples with different capping layers. The laser power was set to 25 mW for each samples. We observe relative demagnetization amplitude $q_{S1}$~40 and $q_{S2}$~50 % for the Pt(3) and the [Co(0.6)/Pt(1.1)]$_4$ capping layer respectively while the relative demagnetization amplitude is much weaker for MgO(5) capping layer and no capping layer ($q_{S3}$~15% and $q_{S4}$~10% respectively). Since the thick Cu layer (100 nm) ensures negligible direct IR excitation and we have excluded demagnetization induced by thermal transport in such samples [19], the demagnetization we observed (figure 2a) is caused by hot-electrons only. In figure 2b, we have plotted the relative demagnetization amplitude q for each samples divided by the demagnetization amplitude observed in the case of the Pt(3) capping layer. It is worth noticing that the relative demagnetization amplitude for this sample and this laser power is consistent with our previous experiments [19]. Interestingly, the ratio between $q_{S4}$ and $q_{S1}$ is in excellent agreement with our previous calculations in which we have estimated that 80% of the hot-electrons were produced within the Pt(3) layer [19]. We wonder about the slightly larger relative demagnetization amplitude for the sample with MgO(5) capping layer compared to the sample without capping layer on top of the Cu(100). We suppose that defects, such as oxygen vacancies, could be at the origin of laser absorption and hot-electron conduction in the MgO layer [28]. The predominance of the Pt(3) capping layer over the Cu(100) layer for the production of hot-electrons explains also the ~30% larger demagnetization amplitude reported for sample 2 ([Co(0.6)/Pt(1.1)]$_4$ capping layer) than for sample 1 (Pt(3) capping layer). The thickness of the former layer is ~7.4 nm while it is only 3 nm in the latter. However, to sustain this assumption, we have investigated the dependence of q on the Pt thickness.

In figure 3, we show the transient Kerr rotation upon hot-electron excitations as well as the maximum relative demagnetization amplitude q as a function of the thickness of the Pt capping layer. The laser power was reduced to 10 mW for this set of experiments. For x=3nm, we obtained q=12,3 % which is rather close to the value we have previously reported in a similar sample for comparable laser power [19]. The relative demagnetization amplitude increases with Pt thickness to reach an extremum (q=21.7 %) for $x_{max}$=7 nm. Then, q decreases for larger Pt thicknesses. The increase of q between x = 3 and x = 7 nm is explained by the much larger CE for Pt than for Cu (figure 2b). Since the typical penetration depth of IR laser pulses is ~10 nm in metals [29], by increasing the Pt thickness, we increase the Pt content in the laser excited top layer of our sample which increases the production of hot-electrons. We remark that the relative demagnetization amplitude is increased by ~45 % for Pt(7) compare to Pt(3). We have shown in figure 2a an increase of ~30% for the [Co(0.6)/Pt(1.1)]$_4$ layer as compared to Pt(3). Therefore, the [Co(0.6)/Pt(1.1)]$_4$ layer, whose total thickness is ~7 nm, appears less efficient than the Pt layer for the production of hot-electrons at equivalent thickness. Spontaneously, we propose that diffusion at the numerous interfaces in the [Co(0.6)/Pt(1.1)]$_4$ layer as well as spin dependent scattering explain the lower demagnetization. We assume that the decrease above x = 7 nm is caused by the much shorter electron lifetime in Pt than in Cu. For thicker Pt layers, the hot-electrons generated on the topmost layer of the Pt(x)/Cu(100) capping are scattered in the Pt layer.

To sustain our scenario, we have performed calculations based on the superdiffusive spin transport model [23, 24]. The simulations were thus run for FM(4)/Cu(100)/Pt(x) multilayers, where FM is a uniform ferromagnetic layer and Pt thickness, x, was varied in the range from 1 to 20 nm. Notably, our model cannot yet accurately describe the hot electron propagation in the Co/Pt multilayers. One can expect that the presence of multiple interfaces can modify the current induced demagnetization amplitude. However, this should have no effect on the dependence on the Pt thickness. We have assumed that the laser pulse excites electrons in the Pt(x) layer uniformly up to 1.5 eV above the Fermi level. The initial excited density of hot electrons is unpolarized. The number of excited electrons in the Pt layer is determined by the laser fluence 2.8 mJ/cm$^2$ and its spatial distribution depends on the distance from the top interface of the multilayer, as $N(x) = N_0 \exp(-x/\lambda_{Pt})$, where $N_0$ is the number of electrons excited at the interface, and $\lambda_{Pt.}$ is the IR penetration depth in Pt. Moreover, in the simulations we have assumed a 40 fs Gaussian laser pulse to match the experimental conditions. The energy and spin-dependent electron velocities and life times were obtained from ab initio calculations [30].

In figure 3b, we report the theoretical calculations for two different values of the light penetration depth in Pt ($\lambda_{Pt}$). Theoretical calculations predict a maximum demagnetization for 8.5 nm of Pt for $\lambda_{Pt} = 11$ nm and 7 nm of Pt for $\lambda_{Pt} = 9$ nm. Then, the best matching of numerical simulations and experimental data occurs for $\lambda_{Pt} = 9$ nm, which is a consistent value [31]. An important point is that the electron lifetime in Cu is rather large compared to Pt. For electron energies less than 0.7 eV above the Fermi level it is already more than 100fs, which translates to mean free paths above 40nm [32]. Together with the weaker optical absorption in Cu this explains the prevalence of the Pt capping layer for the hot-electron induced demagnetization.

**Conclusions**

We have shown that the photo-excitation of hot-electrons is more efficient in Pt(x)/Cu(100) and [Co(0.6)/Pt(1.1)]$_4$/Cu(100) capping layers than in Cu(100) or MgO(5)/Cu(100) capping layers. The order of magnitude for hot-electron production between the Pt(3)/Cu(100) and Cu(100) capping layers is consistent with our previous calculations [19]. Both the experiment and theory has shown that the demagnetization amplitude reach a maximum for a Pt thickness of 7 nm which defines the best compromise between the laser absorption depth and the scattering of hot-electrons in the Pt layer. Our experimental results are sustained by numerical simulations based on the superdiffusive spin transport models. They also provide valuable information to engineer optimized capping layers for the generation of ultrashort hot-electron pulses.

**Acknowledgments**

This work was supported partly by the French PIA project "Lorraine Université d'Excellence", reference ANR-15-IDEX-04-LUE, by the Project Plus cofounder by the "FEDER-FSE Lorraine et Massif Vosges 2014-2020", a European Union Program, and by the OVNI project from Region Grand-Est and by the MATELAS projet institut Carnot ICEEL. K.C. and P. B. acknowledge support from the Czech Science Foundation (Grant No. 18-07172S).

The data that support the findings of this study are available from the corresponding author upon reasonable request.

**Figures:**

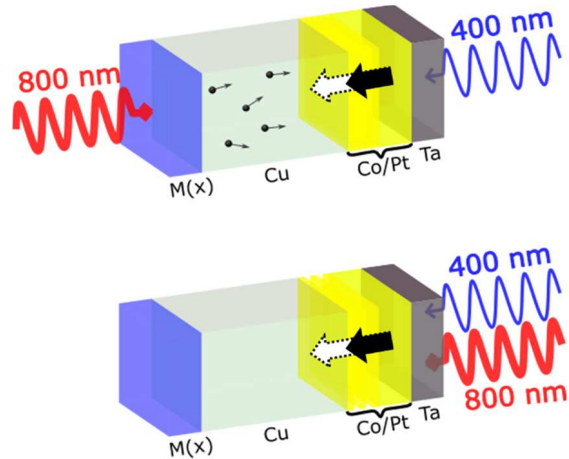

*Figure 1: Schemes of the two experimental configurations for the pump (800 nm) and the probe (400 nm) pulses we have used to record the time-resolved MOKE signals: the "pump-front / probe-back" (top) and "pump-back / probe-back" (bottom). The empty and filled arrows depict the magnetization of the Co/Pt multilayer at equilibrium or in the excited states respectively.*

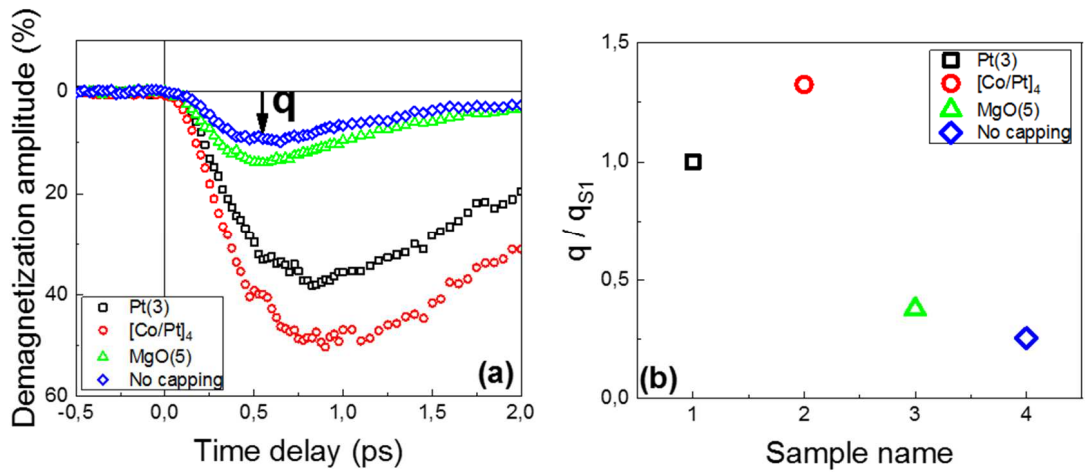

*Figure 2: (a) Transient Kerr rotation upon hot-electrons excitation as a function of the pump probe delay for Pt(3) (black square), [Co(0.6)/Pt(1.1)]$_4$ (red circle), MgO(5) (green triangle) and no capping (blue lozenge). (b) Hot-electrons induced maximum demagnetization amplitude (q) The data are normalized by the q value measured in case of Pt(3).*

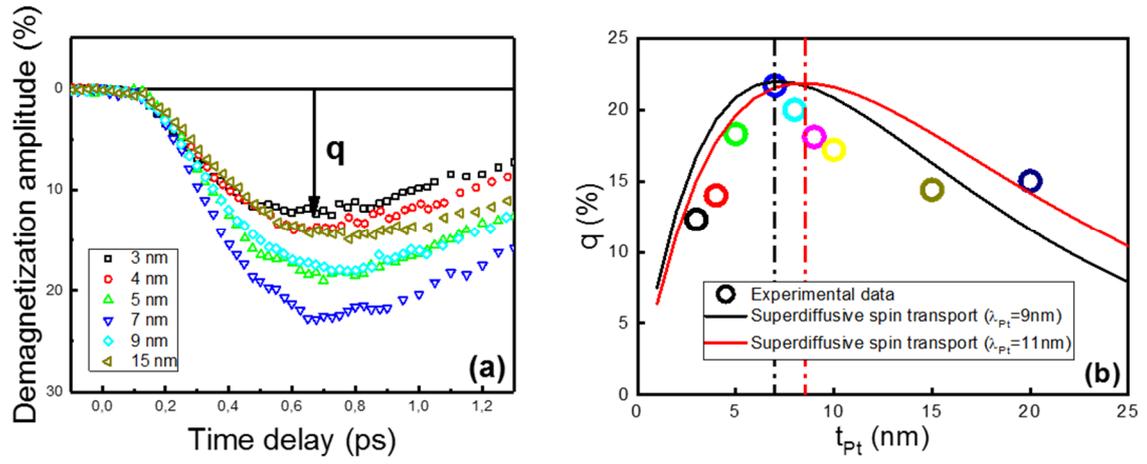

*Figure 3: (a) Transient Kerr rotation upon hot-electrons excitation as a function of the delay for different thicknesses of the Pt top layer. (b) Hot-electrons induced maximum demagnetization amplitude (q) as a function of the thickness of the Pt top layer.*